\documentclass[apjl]{emulateapj}
\usepackage{rotate}
\usepackage{amsmath}
\usepackage{natbib}
\bibpunct{(}{)}{;}{a}{}{,}

\usepackage{graphicx,psfig,epsf}  
\usepackage{times}
\usepackage{subfigure}
\usepackage{textcomp}
\usepackage{multirow}
\usepackage{amsmath}

\def\apjl{ApJ}
\def\apjl{ApJ}                
\def\apjs{ApJS}               
\def\ao{Appl.~Opt.}           
\def\aap{A\&A}                

\def\xmm {\emph{XMM-Newton}}

\received{}
\revised{}
\accepted{}

\begin{document}

\shorttitle{The ULXs in the galaxy NGC 7456}
\shortauthors{Pintore et al.}

\author{F.\,Pintore\altaffilmark{1},
M.\,Marelli\altaffilmark{1},
R.\,Salvaterra\altaffilmark{1},
G.\,L.\,Israel\altaffilmark{2},
G.\,A.\,Rodr\'iguez Castillo\altaffilmark{2},
P.\,Esposito\altaffilmark{3,1}, \\ 
A.\,Belfiore\altaffilmark{1} , 
A.\,De Luca\altaffilmark{1,4},
A.\,Wolter\altaffilmark{5}, 
S.\,Mereghetti\altaffilmark{1},
L.\,Stella\altaffilmark{2},
M.\,Rigoselli\altaffilmark{1,6},
H.\,P.\,Earnshaw\altaffilmark{7},
C.\,Pinto\altaffilmark{8,9}, \\
T.\,P.\,Roberts\altaffilmark{10},
D.\,J.\,Walton\altaffilmark{11},
F.\,Bernardini\altaffilmark{2,12,13},
F.\,Haberl\altaffilmark{14},
C.\,Salvaggio\altaffilmark{5,6},
A.\,Tiengo\altaffilmark{3,1,4},
L.\,Zampieri\altaffilmark{15},
M.\,Bachetti\altaffilmark{16},
M.\,Brightman\altaffilmark{7},
P.\,Casella\altaffilmark{2},
D.\,D'Agostino\altaffilmark{17}, S.\,Dall'Osso\altaffilmark{18},
F.\,F\"urst\altaffilmark{19},  F.\,A.\,Harrison\altaffilmark{7}, M.\,Mapelli\altaffilmark{14,20,21,22}, 
A.\,Papitto\altaffilmark{2},
and M.\,Middleton\altaffilmark{23}} 

\altaffiltext{1}{INAF--IASF Milano, via A. Corti 12, I-20133 Milano, Italy; email: fabio.pintore@inaf.it}
\altaffiltext{2}{INAF--Osservatorio Astronomico di Roma, via Frascati 33, 00078 Monteporzio Catone, Italy}
\altaffiltext{3}{Scuola Universitaria Superiore IUSS Pavia, piazza della Vittoria 15, 27100 Pavia, Italy}
\altaffiltext{4}{Istituto Nazionale di Fisica Nucleare (INFN), Sezione di Pavia, via A. Bassi 6, 27100 Pavia, Italy}
\altaffiltext{5}{Osservatorio Astronomico di Brera, INAF, via Brera 28, 20121 Milano, Italy}
\altaffiltext{6}{Dipartimento di Fisica G. Occhialini, Universit\`a degli Studi di Milano Bicocca, Piazza della Scienza 3, 20126 Milano, Italy}
\altaffiltext{7}{Cahill Center for Astronomy and Astrophysics, California Institute of Technology, 1216 East California Boulevard, Pasadena, CA 91125, USA}
\altaffiltext{8}{European Space Research and Technology Centre (ESTEC), ESA, Keplerlaan 1, 2201\,AZ Noordwijk, The Netherlands}
\altaffiltext{9}{INAF--IASF Palermo, Via U. La Malfa 153, I-90146 Palermo, Italy}
\altaffiltext{10}{Centre for Extragalactic Astronomy, Department of Physics, Durham University, South Road, Durham DH1 3LE, UK}
\altaffiltext{11}{Institute of Astronomy, Science Operations Department, University of Cambridge, Madingley Road, Cambridge CB3 0HA, UK}
\altaffiltext{12}{INAF--Osservatorio Astronomico di Capodimonte, salita Moiariello 16, 80131 Napoli, Italy}
\altaffiltext{13}{New York University Abu Dhabi, Saadiyat Island, PO Box 129188 Abu Dhabi, United Arab Emirates}
\altaffiltext{14}{Max-Planck-Institut f{\"u}r extraterrestrische Physik, Gie{\ss}enbachstra{\ss}e 1, 85748 Garching, Germany}
\altaffiltext{15}{INAF--Osservatorio Astronomico di Padova, vicolo dell'Osservatorio 5, 35122 Padova, Italy}
\altaffiltext{16}{INAF--Osservatorio Astronomico di Cagliari, via della Scienza 5, 09047 Selargius (CA), Italy}
\altaffiltext{17}{Istituto di Matematica Applicata e Tecnologie Informatiche (IMATI) `E. Magenes', CNR, via de Marini 6, 16149 Genova, Italy}
\altaffiltext{18}{Department of Physics and Astronomy, Stony Brook University, Stony Brook 11579 NY, USA}
\altaffiltext{19}{European Space Astronomy Centre (ESAC), ESA, Camino Bajo del Castillo s/n, Villanueva de la Ca\~nada, 28692 Madrid, Spain}
\altaffiltext{20}{Dipartimento di Fisica e Astronomia `Galileo Galilei', Universit\`a di Padova, via F. Marzolo 8, 35131 Padova, Italy}
\altaffiltext{21}{Institut f\"ur Astro- und Teilchenphysik, Universit\"at Innsbruck, Technikerstrasse 25/8, 6020, Innsbruck, Austria}
\altaffiltext{22}{Istituto Nazionale di Fisica Nucleare (INFN), Sezione di Padova, via F. Marzolo 8, 35131 Padova, Italy}
\altaffiltext{23}{Department of Physics and Astronomy, University of Southampton, Highfield, Southampton SO17 1BJ, UK}

\title{The Ultraluminous X-ray sources population of the galaxy NGC 7456}

\begin{abstract}
Ultraluminous X-ray sources (ULXs) are a class of accreting compact objects with X-ray luminosities above $10^{39}$ erg s$^{-1}$. The ULX population counts several hundreds objects but only a minor fraction is well studied. Here we present a detailed analysis of all ULXs hosted in the galaxy NGC 7456. It was observed in X-rays only once in the past (in 2005) by \xmm\, but the observation was short and strongly affected by high background. In 2018, we obtained a new, deeper ($\sim90$ ks) \xmm\ observation that allowed us to perform a detailed characterization of the ULXs hosted in the galaxy. ULX-1 and ULX-2, the two brightest objects ($L_X\sim6-10\times10^{39}$ erg s$^{-1}$), have spectra that can be described by a two-thermal component model as often found in ULXs. ULX-1 shows also one order of magnitude in flux variability on short-term timescales (hundreds to thousand ks). The other sources (ULX-3 and ULX-4) show flux changes of at least an order of magnitude, and these objects may be candidate transient ULXs although longer X-ray monitoring or further studies are required to ascribe them to the ULX population. In addition, we found a previously undetected source that might be a new candidate ULX (labelled as ULX-5) with a luminosity of $\sim10^{39}$ erg s$^{-1}$ and hard power-law spectral shape, whose nature is still unclear and for which a background Active Galactic Nucleus cannot be excluded. 
We discuss the properties of all the ULXs in NGC 7456 within the framework of super-Eddington accretion onto stellar mass compact objects. Although no pulsations were detected, we cannot exclude that the sources host neutron stars.
\end{abstract}


\section{Introduction}

Ultraluminous X-ray sources (ULXs) are a class of accreting compact objects in binary systems, characterized by an assumed isotropic luminosities $L_X>10^{39}$ erg s$^{-1}$ \citep[e.g.][]{fabbiano89,fengsoria11,kaaret17}, i.e. about the Eddington limit for spherical hydrogen accretion onto a 10 $M_{\odot}$ black hole (BH). The ULX population consists of several hundreds sources in nearby galaxies \citep[e.g.][]{earnshaw19}, and they are usually extragalactic (although see \citealt{wilsonhodge18}), off-nuclear and point-like objects. Nowadays, they are believed to be mostly stellar mass BHs or neutron stars (NSs) accreting well above the Eddington limit rather than sub-Eddington accreting intermediate mass BHs ($10^2-10^5$ M$_{\odot}$; e.g. \citealt{colbert99, farrell09}). ULXs strongly challenge our understanding of the accretion processes, especially after the discovery of at least six pulsating ULXs (PULXs) in M82\,X--2 \citep{bachetti13}, NGC 5907 ULX-1 \citep{israel16a}, NGC 7793 P13 \citep{israel16b,fuerst16a}, NGC 300 ULX-1 \citep{carpano18}, M51 ULX-7 \citep{rodriguez19} and NGC 1313 X-2 \citep{sathyaprakash19}, plus the NS candidate in M51 ULX-8 showing a transient cyclotron line \citep{brightman18}. It is now matter of debate whether the ULX populations host preferentially NSs more frequently than BHs  \citep[e.g.][]{middleton17,wiktorowicz17}. It has been proposed that the number of ULXs hosting NSs can be significantly higher than the observed one \citep{king09,pintore17,koliopanos17,walton18}. On the other hand, the detection of $\sim{}30$ M$_\odot$ BHs by LIGO and Virgo \citep[e.g.][]{abbott16,abbott19} has revived the possibility that some ULXs are powered by $>20$ M$_\odot$ BHs \citep[e.g.][]{mapelli09,zampieri09,mapelli10,mapelli13}.

\begin{figure*}
\center
        \includegraphics[width=11.3cm]{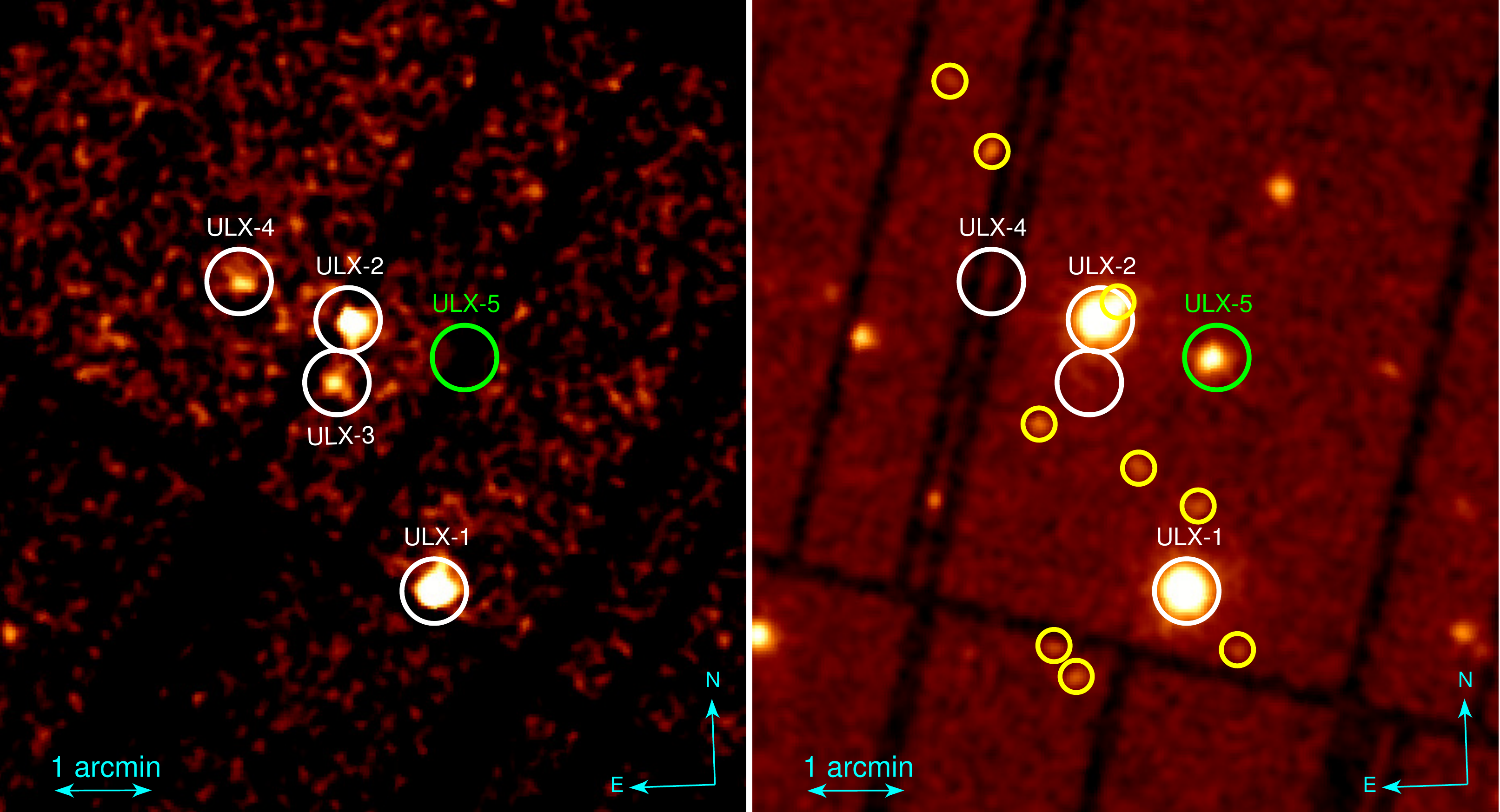}
		\includegraphics[width=6.cm]{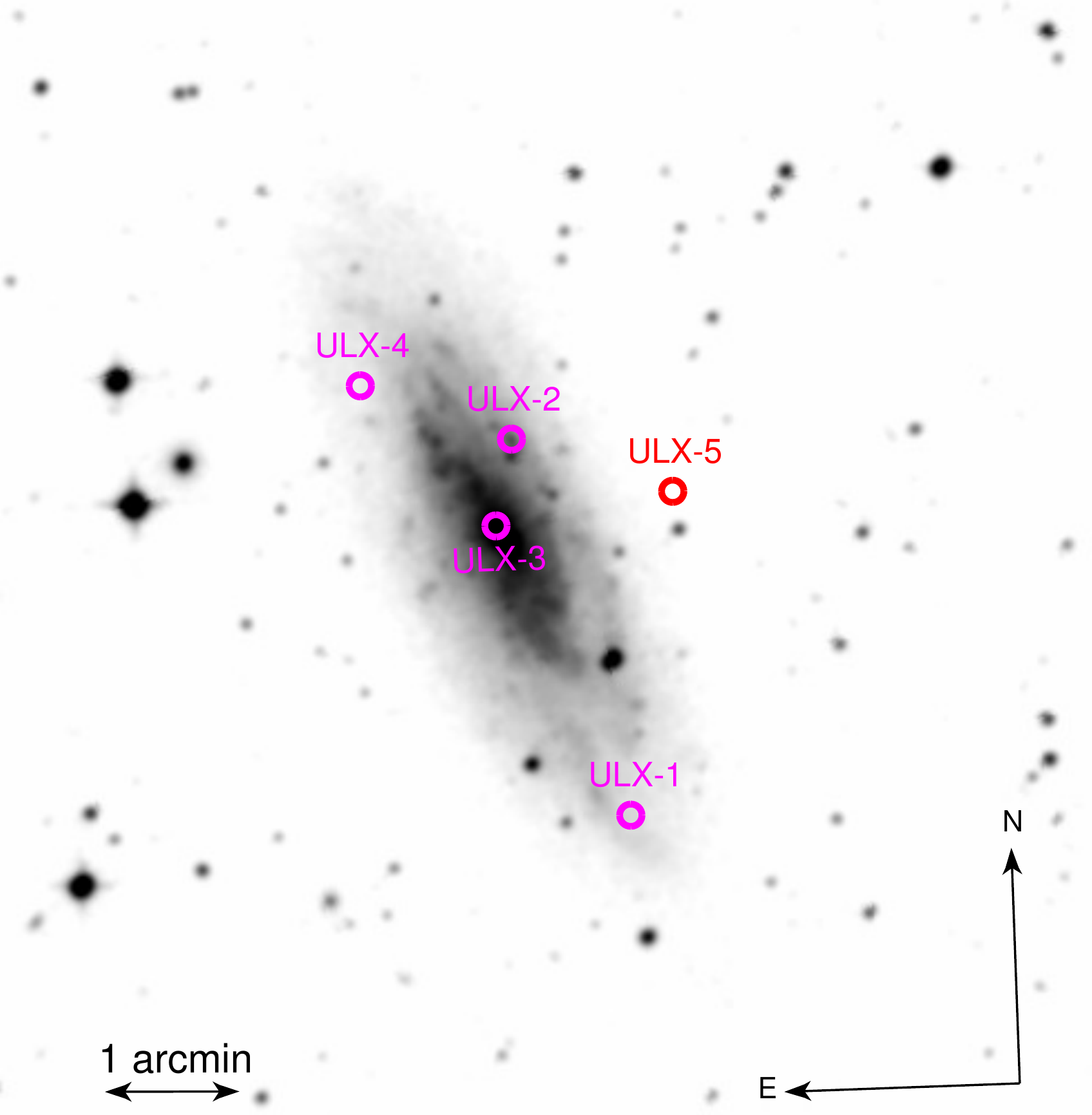}
   \caption{Comparison of the mosaics created by stacking cleaned EPIC pn and MOS1-2 images in the 2005 ({\it left}) and 2018 ({\it center}) \xmm\ observations. Four ULXs (ULX-1, ULX-2, ULX-3 and ULX-4; white circles) are observed during the 2005 observation, while ULX-3 and ULX-4 are not detected anymore in 2018 despite a much longer exposure time. A new source (tentatively labeled as ULX-5, green circle) is detected in the latest observation.  We note that ULX-4 and ULX-5 were not on bad CCD stripes or gaps in the MOS images of 2005 and 2018. Other fainter sources were detected in the galaxy (yellow circles). {\it Right}: DSS optical image of the galaxy NGC 7456. The ULX positions are indicated with magenta circles (radius of 5''), while the ULX-5 is indicated by a red circle.}  
   \label{opt_img}
\end{figure*}

In the case of super-Eddington accretion, it is expected that the accretion disc can eject radiatively powerful outflows/winds \citep[e.g.][]{poutanen07}, which according to magneto-hydrodynamical simulations should be turbulent \citep[e.g.][]{takeuchi13}. The photosphere of these winds may be associated with the soft, thermal component (temperatures of $\sim$0.1--0.5 keV) usually observed at low energies in the ULX spectra \citep{gladstone09}. In addition, the spectra are often accompanied by a hard, thermal-like component with a roll-over at 3--7 keV \citep[e.g.][]{gladstone09}, often interpreted either as an optically-thick corona around the compact object or as the inner regions of a geometrically thick accretion disc \citep[e.g.][]{kawaguchi03}. When {\it NuSTAR} spectra of high signal-to-noise above 10 keV are available, the vast majority of ULXs are also characterized by a third hard component, phenomenologically described with a cut-off power-law model, dominant in the case of the PULXs and, for this reason, possibly associated with the emission of the accretion column above the NS magnetic polar caps \citep[e.g.][]{walton18}.  

The short-term temporal variability properties of the ULXs differ significantly from source to source, but also from observation to observation of the same source. It has been shown that the most variable sources are generally those with the softest spectra \citep[e.g.][]{sutton13}. \citet{earnshaw19} examined the 3XMM-DR4 catalogue, which contains \xmm\ observations performed before 2012, finding that the number of ULXs with high intra-observation variability is however quite limited (8 sources) if compared to the whole population number. In addition, \citet{heil09} showed that the absence of high short-term variability cannot be only due to poor signal-to-noise ratios. The variability seen in the ULXs has been proposed to be associated to the high-energy components which may (extrinsically) vary due to the turbulences of the winds that, from time to time, encounter our line of sight \citep[e.g.][]{middleton15a}. Observational evidence of winds in ULXs has been obtained from the detection of absorption lines with blueshifts of $\sim0.2c$ in high quality \xmm/RGS  spectra of some ULXs \citep[e.g.][]{pinto16}, the discovery of the first extended X-ray bubble around the PULX NGC 5907 ULX-1 \citep{belfiore19} and the optical bubbles around several ULXs \citep[e.g.][]{pakull02}.

Furthermore, significant long-term (days to years) flux variability is observed in the vast majority of the ULXs for which multiple observations are available. However, only a small fraction of ULXs can be considered transient (i.e. with flux variation up to two orders of magnitude). 
Amongst the transient ULXs, the largest variations (up to a factor of 500) have been observed in the PULXs. A possibility to explain this temporal behaviour may be ascribed to the onset/offset of the propeller mechanism \citep[e.g.][]{tsygankov16, israel16a}.
Hence, the transient ULXs clearly play an important role in the ULX scenario as they may be the ideal places to look for new PULX candidates \citep[e.g.][]{earnshaw18,song19}.

Here, we focus on the ULXs populating NGC 7456, a spiral galaxy at a distance of $\sim$15.7 Mpc (\citealt{tully16}; Galactic column density of neutral hydrogen expected along the line of sight of $8.7\times10^{19}$ cm$^{-2}$, \citealt{hi4pi16}), and hosting at least four ULXs \citep{walton11}. The galaxy was observed { for the first time} in the X-rays in 2005 by \xmm, but the observation was short and strongly affected by high background, hence preventing any reliable investigation of the ULX properties. { No other X-ray observations with \xmm\ or other X-ray satellites have been performed since then, until, in 2018, our group obtained a new, deeper observation in the context of an \xmm\ Large Programme (PI: G. Israel).} This new dataset allowed us to perform a more constraining characterization of the ULXs in the galaxy and the probable identification of a new ULX in its outskirts.

\section{Data reduction}

NGC 7456 was observed twice by \xmm, first in May 2005 (Obs.ID: 0303560701; { PI: D. Rosa Gonz\'alez}) and then in May 2018 (Obs.ID: 0824450401; { PI: G. Israel}), with total exposure times of $\sim$$10$ ks and $\sim$$92$ ks, respectively (Table\,\ref{log_obs}). 
The observations were processed with the Scientific Analysis Software (SAS) v.16.1. We reduced the data from the EPIC cameras (all operated in full frame mode and with thin filter), selecting events with FLAG=0, and PATTERN$\leq$4 and PATTERN$\leq$12 for pn and MOS, respectively.
We filtered for high background time intervals only the 2018 observation; indeed the first observation was entirely taken during a high-background period, hence we chose not to filter it for proton-flares in order to avoid rejecting all events of this observation. The net exposures for the 2005 and 2018 observations were $\sim10$ ks and $\sim80$ ks, respectively. 
For pn and MOS data, we extracted source and background events from circular regions of radii of 30'' and 65'', respectively. The photon times of arrival (ToAs) were converted to the Solar System barycenter with the SAS task {\sc barycen}, using the best X-ray coordinates of each source reported in \cite{walton11}.

All the spectra were rebinned with at least 25 counts per bin using the Ftool {\sc grppha}.
In our spectral analysis, we fitted simultaneously the EPIC-pn and EPIC-MOS data, in the 0.3--10 keV range, with {\sc xspec} v.12.10.1 \citep{arnaud96}. In all fits, we included a multiplicative constant to take into account possible mis-calibration of the relative flux between the three instruments. These did not vary for more than 10$\%$, as expected \citep{madsen15}.

We produced background subtracted light curves in the 0.2-12 keV energy band for all sources in the two observations using the EXTraS\footnote{\url{http://www.extras-fp7.eu}} tools and prescriptions (\citealt{deluca16} and De Luca et al. in preparation). Count-rates of light curves from different EPIC cameras were converted into fluxes by using the conversion factor in the Processing Pipeline Subsystem products (PPS) files and then combined \citep{marelli17} to obtain the total light-curve. 

\begin{figure*}
\center
		\includegraphics[width=4.6cm,angle=270]{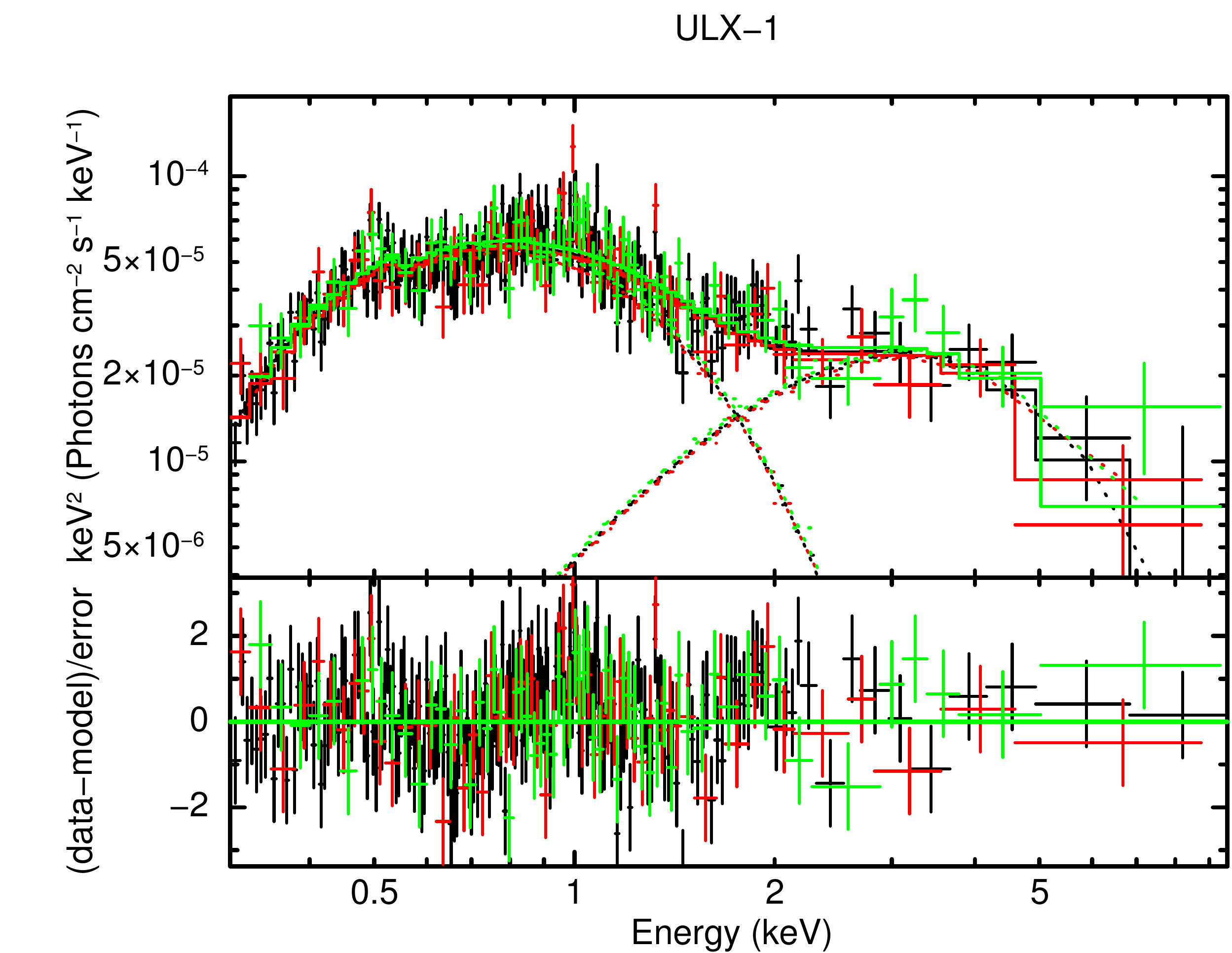}
		\includegraphics[width=4.6cm,angle=270]{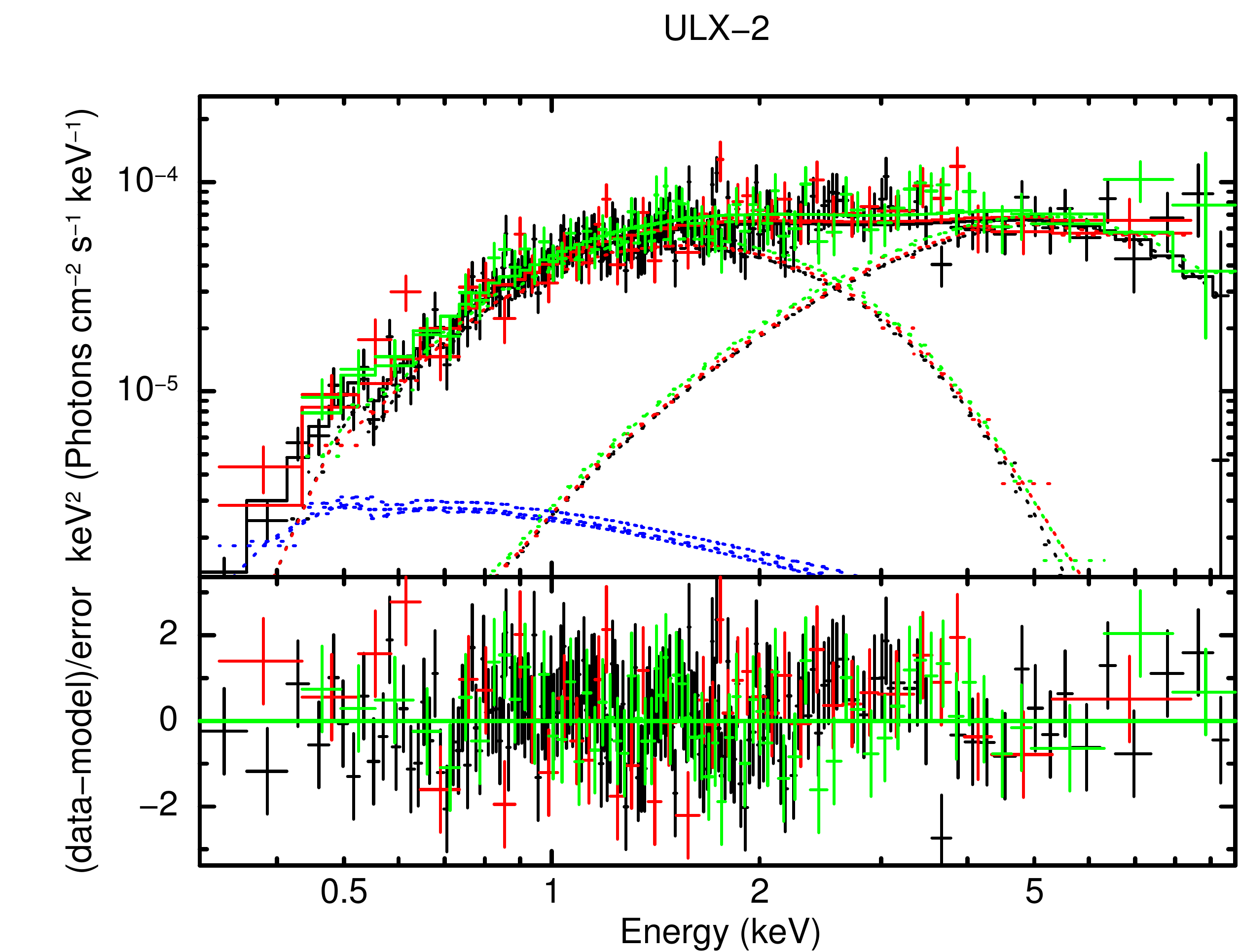}
		\includegraphics[width=4.6cm,angle=270]{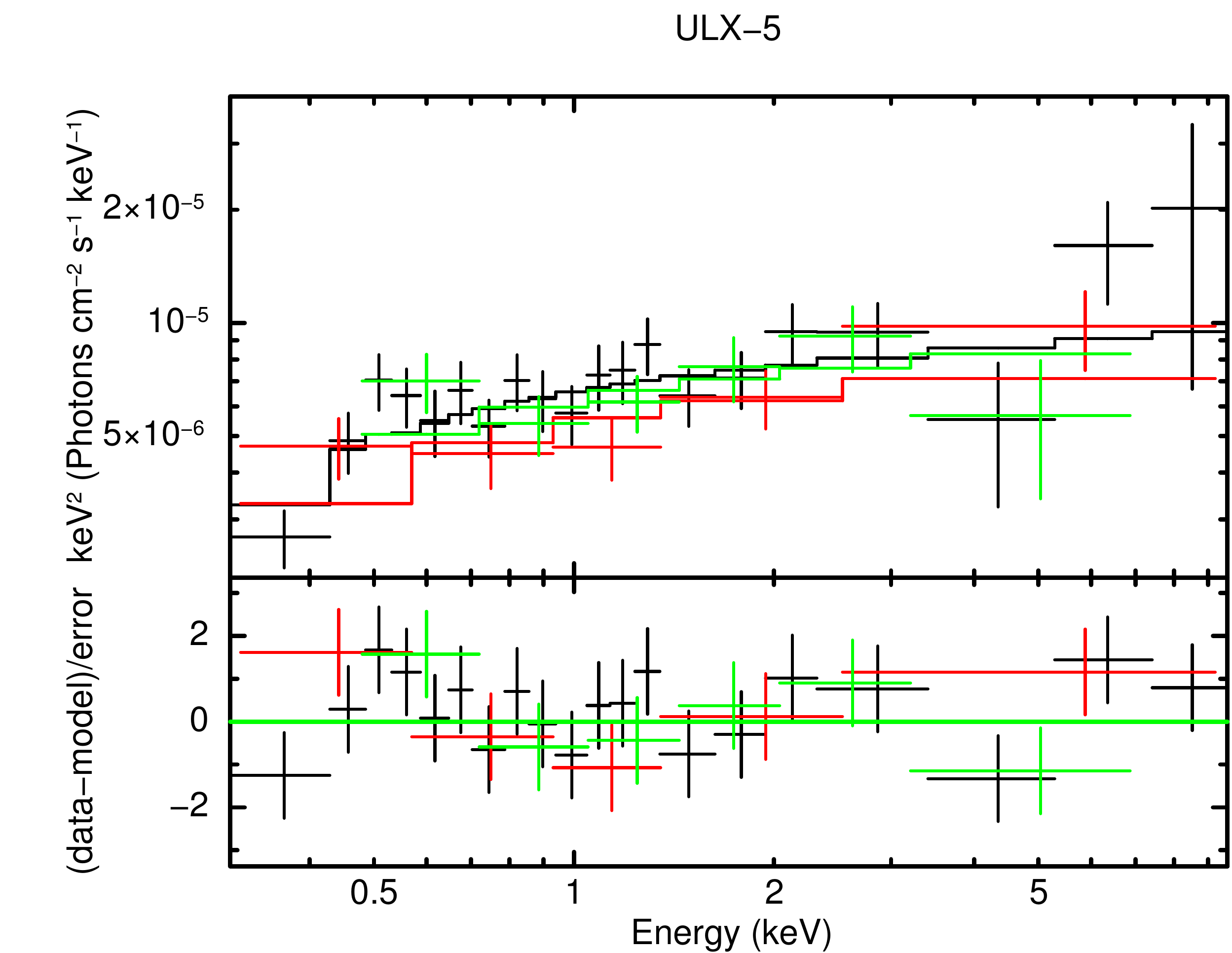}
\caption{ Top: unfolded ($E^2f(E)$) spectra of the ULX-1 ({\sc tbabs(diskbb+bbodyrad)}, left), ULX-2 ({\sc tbabs(diskbb+bbodyrad)}, center) and ULX-5 ({\sc tbabs(powerlaw)}, right). EPIC-pn spectra are indicated with black points, while EPIC-MOS1 and 2 in red and green, respectively. The solid lines represent the best-fit models, while in the bottom panels the corresponding residuals are shown. In the central plot, we also show in blue dashed lines the spectrum of the contaminant source (see Section~\ref{sect_ulx2}).
All spectra have been further rebinned for display purposes only.}  
   \label{spec_ulx}
\end{figure*}

\begin{table}
\caption{Log of the \xmm\ observations.}
\begin{center}
\label{log_obs}
\begin{tabular}{lcccc}
\hline
No. & Obs.ID. & Date & Instrument & Tot. Expos.\\
 &  &  & EPIC-pn &\\
\hline
1& 0303560701 & 2005-05-06 & pn+MOS & 10.2 ks \\ 
2 & 0824450401 & 2018-05-18 & pn+MOS & 92.4 ks \\
\hline
\end{tabular}
\end{center}
\end{table}

\section{Data Analysis}
\citet{walton11} reported the presence of four ULXs in the galaxy NGC 7456 (Figure~\ref{opt_img}-left), using the 2005 \xmm\ observation.
We carried out a source detection on the new 2018 \xmm\ observation, applying the task {\sc edetect\_chain} (setting a likelihood threshold limit to 15) to the combined EPIC-pn and EPIC-MOS images in the 0.2-12 keV energy band. We found that only ULX-1 and ULX-2 were significantly detected. ULX-3 and ULX-4 were instead below the detection threshold, suggesting that they went back to quiescence. 
We also found a previously undetected source (at $\sim78$'', i.e. $\sim6$ kpc of projected distance, from the center of the galaxy), that we tentatively labelled as ULX-5 because it is bright and it lies inside the size of the galaxy NGC 7456 \citep{paturel0}.

\begin{table}
\caption{Positions of the sources found (at $>3\sigma$ significance) in the galaxy NGC 7456 using the 2018 \xmm\ observation. We report the source position, the statistical error and the source count rate (pn plus MOS1+2).}
\begin{center}
\label{pos_ulx}
\begin{tabular}{lcccc}
\hline
Source & RA & Dec. & Stat. err. & Rates$_\text{0.3--10 keV}$ \\
 & & & arcsec & cts s$^{-1}$\\
\hline
ULX-1 & 23:02:05.62 & -39:36:17.0 & 0.1 & 0.159$\pm$0.002\\ 
ULX-2 & 23:02:09.73 & -39:33:26.9 & 0.1 & 0.145$\pm$0.002 \\
ULX-3$^a$ & 23:02:10.64 &  -39:34:04.7 & &  \\
ULX-4$^a$ & 23:02:15.15 & -39:33:01.1 & &  \\
ULX-5 & 23:02:03.80 & -39:33:52.0 & 0.2 & 0.0235$\pm$0.0008\\ 
\hline
X6 & 23:02:08.86 & -39:33:15.2 & 0.3 & 0.0119$\pm$0.0007\\ 
X7 & 23:02:15.30 & -39:31:39.1 & 0.5 & 0.0052$\pm$0.0005\\ 
X8 & 23:02:13.33 & -39:34:29.0 & 0.6 & 0.0039$\pm$0.0004\\ 
X9 & 23:02:11.85 & -39:37:06.1 & 0.7 & 0.0036$\pm$0.0004\\ 
X10 & 23:02:17.41 & -39:30:54.6 & 1.1 & 0.0029$\pm$0.0004\\ 
X11 & 23:02:04.98 & -39:35:23.4 & 0.8 & 0.0029$\pm$0.0004\\ 
X12 & 23:02:08.09 & -39:34:58.6 & 0.9 & 0.0028$\pm$0.0004\\ 
X13 & 23:02:12.98 & -39:36:46.5 & 0.9 & 0.0027$\pm$0.0004\\ 
X14 & 23:02:03.16 & -39:36:53.3 & 1.0 & 0.0018$\pm$0.0003\\ 
\hline
\end{tabular}
\end{center}
\flushleft{$^a$ In 2018, these sources were not detected, hence for completeness we report the coordinates given in \citet{walton11}.}
\end{table}

\begin{table*}
\center
          \caption{Best-fit spectral parameters of the ULXs detected in the \xmm\ observations. Errors are at $90\%$ confidence level for each parameter of interest.} 
\scalebox{0.72}{\begin{minipage}{26cm}
      \label{spec_par}
\begin{tabular}{lcccccccccccccc}
\hline
src. & year &  nH  & $kT_{\mathrm{dbb}}$ & $p$ & Norm. & $kT_{\mathrm{bb}}$/$\Gamma$ & Norm. & E$_{line}$ & $\sigma_{line}$ & N$_{line}$ & Flux$^{*}$ & L$_X$ & $\chi^2_{\nu}/\text{dof}$ \\
 &  & $10^{22}$ cm$^{-2}$ & keV & & & & & keV & keV & keV & $10^{-13}$ erg cm$^{-2}$ s$^{-1}$ & $10^{39}$ erg  s$^{-1}$  & \\
\hline
\multirow{2}{*}{ULX-1} & 2005 & 0.049 (fixed) & $0.29^{+0.04}_{-0.04}$ & - & $2.0^{+1.8}_{-0.9}$ & $1.0^{+0.7}_{-0.4}$ keV & $0.012^{+0.05}_{-0.009}$ & - & - & - & $2.9\pm 0.5$ & $10.0\pm1.0$ & 1.00/84 \\
 & \multirow{2}{*}{2018} & $0.05^{+0.02}_{-0.01}$ & $0.27^{+0.02}_{-0.02}$ & - & $1.8^{+0.7}_{-0.5}$ & $0.80^{+0.09}_{-0.07}$ keV & $0.012^{+0.006}_{-0.004}$ & - & - & - & $1.65\pm0.05$ & $5.9\pm0.4$ & 1.05/354 \\
 & & $0.09^{+0.03}_{-0.03}$ & $0.23^{+0.02}_{-0.03}$ & - & $5.3^{+7.6}_{-2.4}$ & $0.73^{+0.08}_{-0.07}$ keV & $0.019^{+0.01}_{-0.007}$ & $0.67^{+0.03}_{-0.03}$ & $0.09^{+0.06}_{-0.05}$ & $0.07^{+0.1}_{-0.04}$ & $1.63\pm0.05$ & $6.9\pm0.6$ & 0.97/351 \\
\hline
\multirow{2}{*}{ULX-2} & \multirow{2}{*}{2005} &  $0.4^{+0.2}_{-0.2}$ & - & - & - & $2.0^{+0.4}_{-0.3}$ & $(6.4^{+3.2}_{-2.2})\times10^{-5}$ & - & - & - & $2.3_{-0.7}^{+0.6}$ & $6.9\pm0.2$ & 1.33/32 \\
 & &  $0.08^{+0.2}_{-0.08}$ & $1.3^{+0.3}_{-0.3}$ & - & $4^{+5}_{-2}\times10^{-5}$ & - & - & - & - & - & $2.3_{-0.7}^{+0.6}$ & $6.9\pm0.2$ & 1.20/32 \\
& \multirow{2}{*}{2018} &  $0.24^{+0.05}_{-0.04}$ & $0.56^{+0.1}_{-0.09}$ & & $0.09^{+0.09}_{-0.04}$ & $1.3^{+0.4}_{-0.2}$ keV & $(4.2^{+4.3}_{-2.6})\times10^{-3}$ & - & - & - & $2.3\pm0.1$ & $8.4\pm0.2$ & 1.07/311 \\
& &  $0.35^{+0.02}_{-0.03}$ & $3.1^{+0.7}_{-0.5}$ & $0.5^{+0.01}_{0.01}$ & $(2.7^{+3.5}_{-1.4})\times10^{-5}$ & - & - & - & - & - & $2.4\pm0.1$ & $9.9\pm0.2$ & 1.04/312 \\
\hline
ULX-5 & 2018 &  $0.04^{+0.04}_{-0.03}$ & - & - & - & $1.9^{+0.2}_{-0.2}$ & $7.0^{+1.0}_{-0.8}$ & - & - & - & $0.40^{+0.05}_{-0.05}$ & $1.3\pm0.1$ & 0.90/81 \\
\hline

\end{tabular}
\end{minipage}}
\flushleft $^*$ EPIC absorbed flux in the 0.3--10 keV energy band; 
\end{table*}

In addition, super-imposed to the galaxy, we detected at least nine other fainter X-ray sources (Table~\ref{pos_ulx}; yellow circles in Fig.~\ref{opt_img}-center), with fluxes between $\sim$$(0.4-2)\times10^{-14}$ erg cm$^{-2}$ s$^{-1}$. Assuming they are all hosted in NGC 7456 and have an absorbed power-law spectral shape with nH $=10^{21}$ cm$^{-2}$ and $\Gamma=2$, their 0.3--10 keV luminosities are 1--$6\times10^{38}$ erg s$^{-1}$. A fit with a constant model of their lightcurves binned with $\Delta T=10$ ks showed that, except for one, all the sources did not display significant flux-variability during the observation, with null hypothesis probability (NHP) higher than 0.05. 

\subsection{ULX-1}

\subsubsection{Spectral analysis}
We fitted simultaneously the time-averaged EPIC-pn and MOS spectra of the 2018 observation. 
We first adopted a simple phenomenological model, i.e. an absorbed multi-colour blackbody disc ({\sc diskbb}; \citealt{mitsuda84}) plus a blackbody ({\sc bbodyrad} in {\sc xspec}), which, as in other ULXs, provided a good description of the 0.3--10 keV spectra \citep[e.g.][]{stobbart06, pintore15, rodriguez19}. As commonly found with ULXs, we note that other alternative two-component models as a {\sc diskbb+nthcomp}, a {\sc diskbb+cutoffpl} or a {\sc diskbb+highecut$\times$powerlaw} provide good fits as well.

\begin{figure*}
\center
    \includegraphics[width=6.1cm]{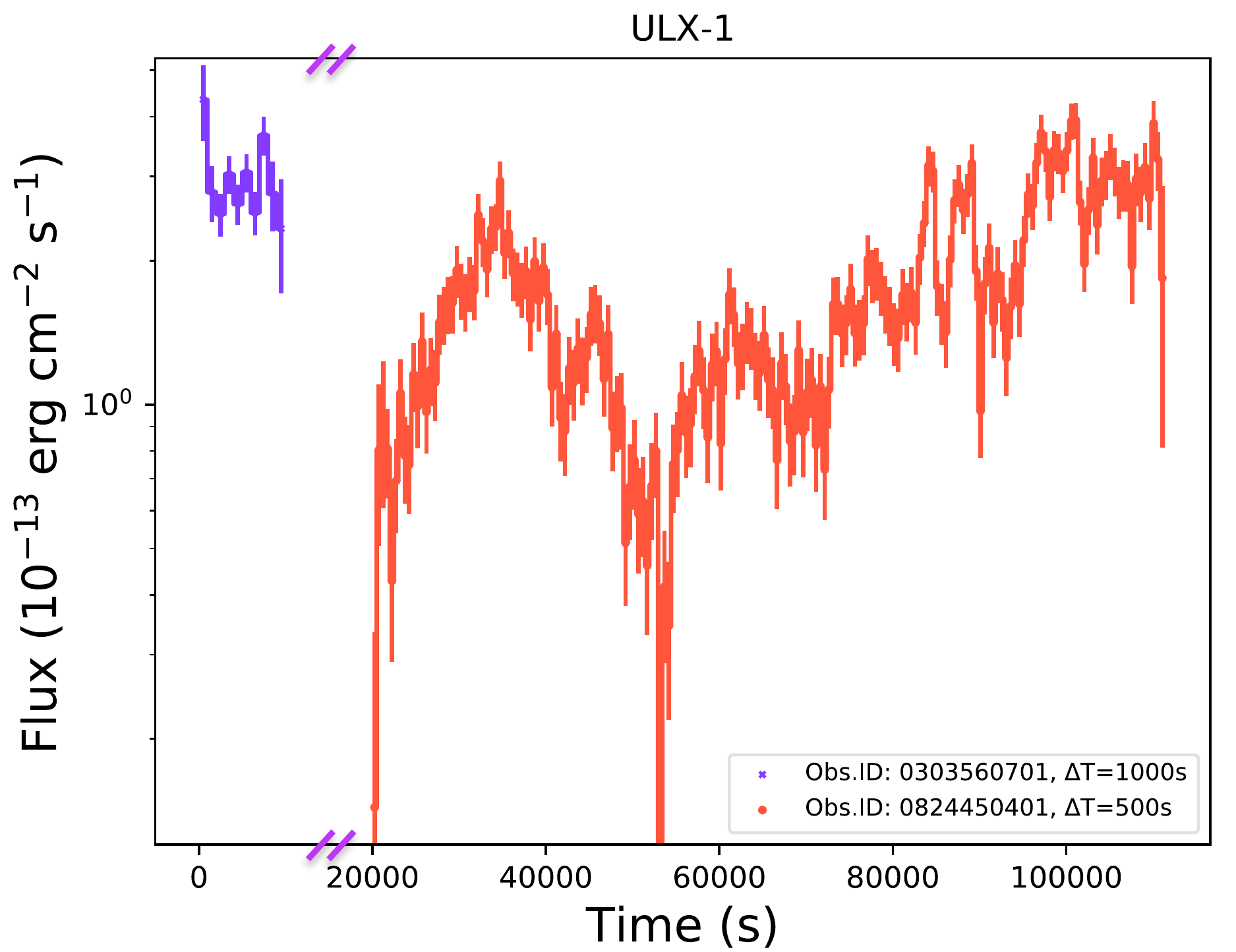}
   \includegraphics[width=5.8cm]{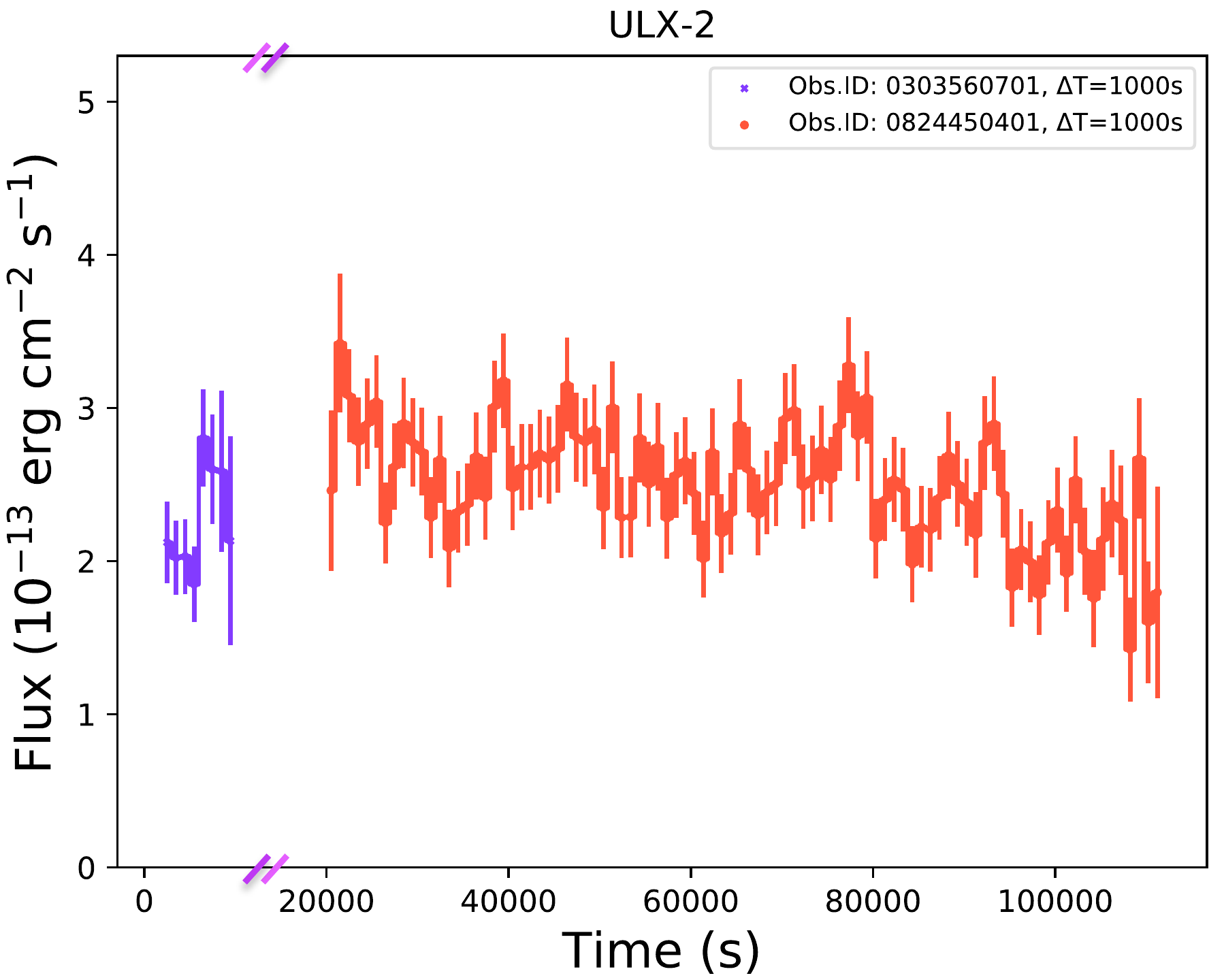}
   \includegraphics[width=5.9cm]{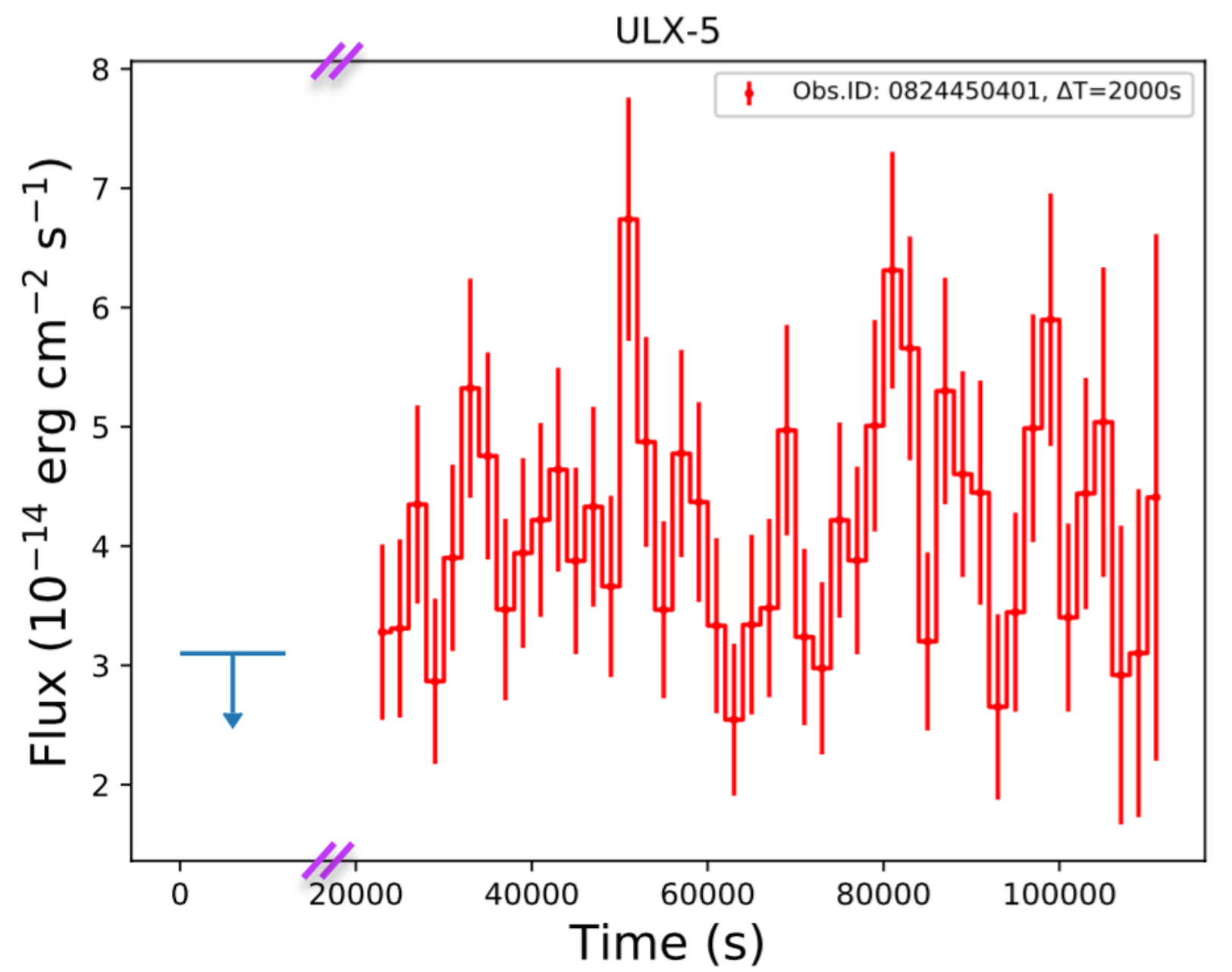}
   \caption{Lightcurves in the 0.2--12 keV energy band of ULX-1 (left), ULX-2 (center) and ULX-5 (right) during the 2005 (blue) and 2018 (red) \xmm\ observations. The X-axis is arbitrary as the gap within the two observations was shortened for displaying purposes.}  
   \label{lc_ulx}
\end{figure*}

The {\sc tbabs(diskbb+bbodyrad)} model gave a good fit ($\chi^2/\text{dof}\sim371.7/354$); we report the best-fit parameters in Table~\ref{spec_par}. We found a column density ($\sim5\times10^{20}$ cm$^{-2}$) higher than the Galactic one, and temperatures of kT$_{\mathrm{dbb}}\sim0.27$ keV and kT$_{\mathrm{bb}}\sim0.8$ keV. The latter are associated to emitting radii of $\sim$2000--7000 km (assuming an unknown inclination angle between 5$^\circ$--85$^\circ$ and no color-correction factor) and $170\pm35$ km for the {\sc diskbb} and {\sc bbodyrad} models, respectively.
ULX-1 was quite soft (Figure~\ref{spec_ulx}-left), with an unabsorbed 0.3--10 keV flux of $(2.0\pm0.1)\times10^{-13}$ erg cm$^{-2}$ s$^{-1}$, implying an unabsorbed luminosity of $(6.0\pm0.4)\times10^{39}$ erg s$^{-1}$.

We note that, in the case of super-Eddington accretion, modeling the accretion disc with an optically thick and geometrically thin disc may not be appropriate. In such a scenario, the disc should be instead modelled with a ``slim'' disc (i.e. a  geometrically and optically thick disc). Some authors \citep[e.g.][]{walton18} showed that a possible description of the ULX spectra is given by the combination of {\sc diskbb} and {\sc diskpbb} models (where, in the latter, the radial dependence of the temperature is given by $r^{-p}$, and $p$ is 0.75 for a standard disc and 0.5 for an advection dominated disc).
Thus, we tentatively substituted firstly the {\sc diskbb} with a {\sc diskpbb} model: this model was suitable for the data as well, although we found that $p$ was unconstrained. Secondly, we also substituted the high energy {\sc bbodyrad} with the {\sc diskpbb} model: even in this case, although the fit was formally good, the $p$ parameter was not constrained. This implies that the current data quality does not allow us to rule out the existence of a thick disc yet in ULX-1. 

Hereafter, we will consider only the {\sc diskbb+bbodyrad} model results. Although the best-fit with this model is statistically acceptable, some residuals in absorption around 0.6--0.7 keV (or, alternatively, in emission around 1 keV), are still observed (see Fig.~\ref{spec_ulx}-left). These are often found in the ULX spectra \citep[see e.g.][]{middleton15b}. We can fit the residual in absorption with a Gaussian absorption line ({\sc gabs} in {\sc xspec}), obtaining an improvement of $\Delta\chi^2=33$ for 3 additional d.o.f. The inclusion of this feature slightly changed the continuum spectral parameters, giving nH $=(9\pm3)\times10^{20}$ cm$^{-2}$, $kT_{\mathrm{dbb}}=0.23^{+0.02}_{-0.03}$ keV and $kT_{\mathrm{bb}}=0.73^{+0.08}_{-0.07}$ keV. The line energy, its full-width at half maximum ($\sigma$) and the line depth (i.e. the equivalent width) converged to $E=0.67\pm0.03$ keV, $\sigma=0.09^{+0.06}_{-0.05}$ keV and $0.07^{+0.1}_{-0.04}$ keV, respectively. 
We do not consider this feature as an artifact of the modeling because the two spectral components intersect around 1.5--2 keV, i.e. well above the line energy. Hence, should the feature be real, it might be associated to a blending of ionized oxygen lines in absorption (O VII-VIII at $\sim0.5-0.7$ keV). ULX-1 was not in the field of view of the RGS instrument in 2018 and its flux was anyway too low for the RGS instruments, therefore no high resolution X-ray spectra are currently available to further investigate the nature of such a possible feature. 

\begin{figure}
\center
   		\includegraphics[width=9.3cm]{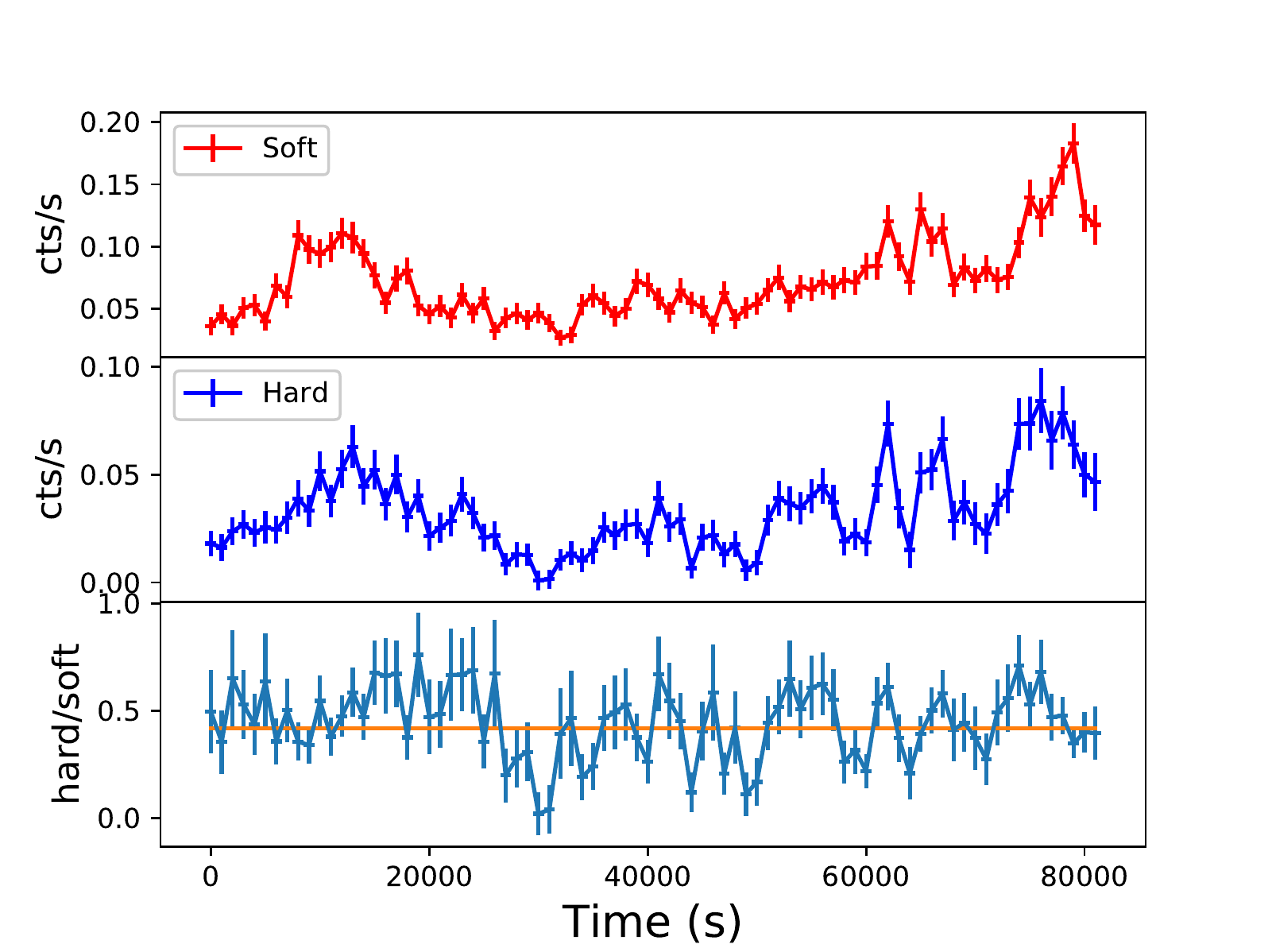}
   		\includegraphics[width=9.3cm]{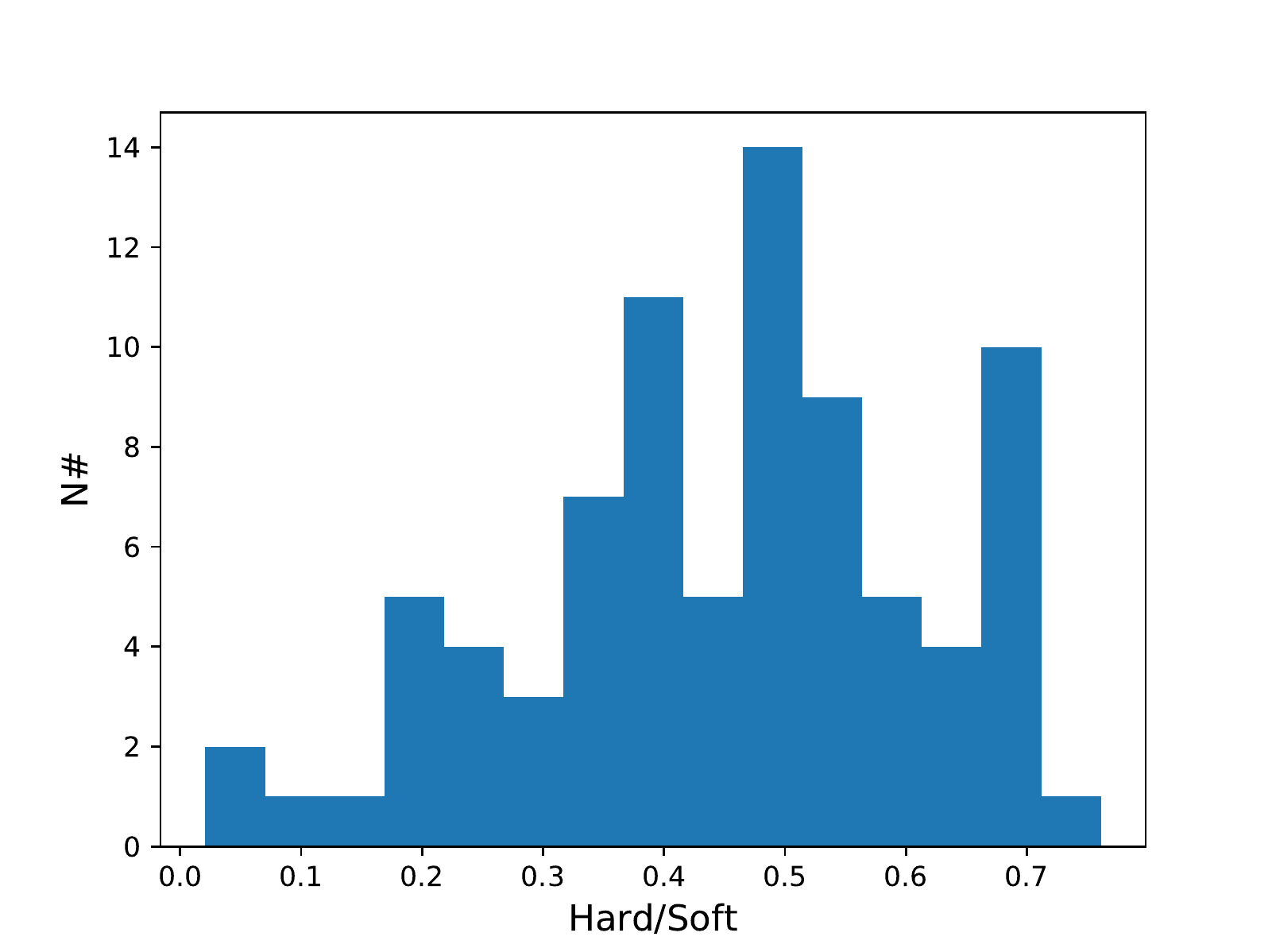}
\caption{Top: EPIC-pn soft (0.3--1.0 keV, top panel) and hard (1.0--10 keV, center panel) background-subtracted lightcurves of ULX-1 ($\Delta T=1000$s) compared with their hardness ratio (bottom panel), during the 2018 \xmm\ observation. The orange line is the best-fit ($H/S=0.42$) with a constant model. Bottom: distribution of the hardness ratios that indicates how the spectral variability is not normally distributed.}  
   \label{lc_hardness}
\end{figure}

Finally, we note that the {\sc tbabs(diskbb+bbodyrad)} model is adequate for the 2005 data as well, although we had to fix the absorption to the 2018 best-fit nH ($4.9\times10^{20}$ cm$^{-2}$), as this is otherwise unconstrained. The other best-fit parameters are shown in Table~\ref{spec_par} and they are consistent with those of 2018 to within uncertainties. On the other hand, we remark that these results have to be considered with caution as the observation was fully affected by high background. 
We measured an unabsorbed 0.3--10 keV flux of $(3.4\pm0.5)\times10^{-13}$ erg cm$^{-2}$ s$^{-1}$ corresponding to a luminosity of $(1.0\pm0.1)\times10^{40}$ erg s$^{-1}$ (in agreement with \citealt{walton11}), a factor of $\sim2$ higher than the average flux in 2018.

\subsubsection{Temporal properties}
In 2005, ULX-1 was the brightest source in NGC 7456 at a luminosity of $\sim10^{40}$ erg s$^{-1}$ \citep{walton11}. Because of the short exposure time and high background level, its 0.3--10 keV \xmm\ light-curve accumulated in bins of 500s (Figure~\ref{lc_ulx}-left) gives little information on the source variability.

Instead, the longer 2018 \xmm\ observation allowed us to find a highly significant variability with recurrent increments of flux up to an order of magnitude (see Figure~\ref{lc_ulx}-left), on time-scales of only a few ks. We found that the 0.3--10 keV fractional variability F$_{var}$ \citep[e.g.][]{vaughan03}, on time-scales $>$500s, is $47\pm1\%$: this is one of the highest non-periodic short-term variability presently measured in a ULX \citep[e.g.][]{sutton13, middleton15a}. We verified that such a high variability is also preserved on smaller time-scales ($<500$s).

To better assess the origin of such a variability, we evaluated the hardness ratios of the net counts between the 0.3--1.0 keV and 1.0--10 keV energy bands, chosen so that in each band the number of counts is comparable.
The hard-to-soft band ratio (H/S) clearly shows a variability during the 80 ks of observation, in particular after the first $\sim$25 ks of the observation where there is an indication of a softening which apparently tracks periods of very low fluxes (Figure~\ref{lc_hardness}-top).

Finally, we also searched for coherent pulsations in the 2018 data. Adopting a generalization of the Fourier-based procedure described in \citet{israel96} (see also \citealt{rodriguez19}), we could only place $3\sigma$ upper limits on the pulse fraction of $17-19\%$ for periods in the range 150ms--500s, assuming a sinusoidal pulse profile. This limit is, on average, slightly larger than the pulse fraction observed in PULXs.

\subsubsection{Hardness-resolved spectroscopy}
In this section, we investigate in greater detail the spectral variability of the 2018 observation suggested by the hardness ratios. Here we perform hardness-resolved spectroscopy to track or identify specific patterns with the ULX spectral state. Figure~\ref{lc_hardness}-bottom suggests the existence of at least two different spectral states, approximately above and below $H/S=0.42$. We selected this threshold to extract two EPIC spectra for both states, and we fitted them simultaneously, adopting the same model of the averaged spectrum (i.e. {\sc tbabs(diskbb+bbodyrad)}). 
As a first step, we left all the parameters free to vary independently. We found that the normalizations of the soft component and the the temperatures of the two thermal components were consistent, to within uncertainties, between the two spectra and for this reason we kept them linked. Instead the hottest blackbody normalizations was different for the two spectra, indicating that the flux variations are mostly driven by the high energy component. This fit gave $\chi^2/\text{dof}=420/352$. 

However, several residuals around 0.7 keV were still present, as already seen for the average spectrum. Also in this case, we added a {\sc gabs} model, fixing the feature energy and width (0.67 keV and 0.09 keV) to the best-fit values found for the average spectrum. The final best-fit ($\chi^2/\text{dof}=389.15/350$) provided the following parameters: nH $=9^{+3}_{-2}\times10^{20}$ cm$^{-2}$, kT$_{\mathrm{dbb}}=0.22^{+0.02}_{-0.02}$ keV, and kT$_{\mathrm{bb}}=0.68^{+0.06}_{-0.05}$ keV. These are slightly different from those inferred in the average spectrum.
The blackbody normalizations for the low and high $H/S$ spectra were $0.015^{+0.008}_{-0.005}$ ph cm$^{-2}$ s$^{-1}$ and $0.032^{+0.01}_{-0.01}$ ph cm$^{-2}$ s$^{-1}$, respectively, implying that the emitting radii varied from $\sim200$ to $\sim300$ km. The depth of the {\sc gabs} is instead consistent between low and high state at a value of $0.07^{-0.02}_{+0.02}$ keV. 

\subsection{ULX-2}
\label{sect_ulx2}

\begin{figure}
\begin{center}
		\includegraphics[width=8.4cm]{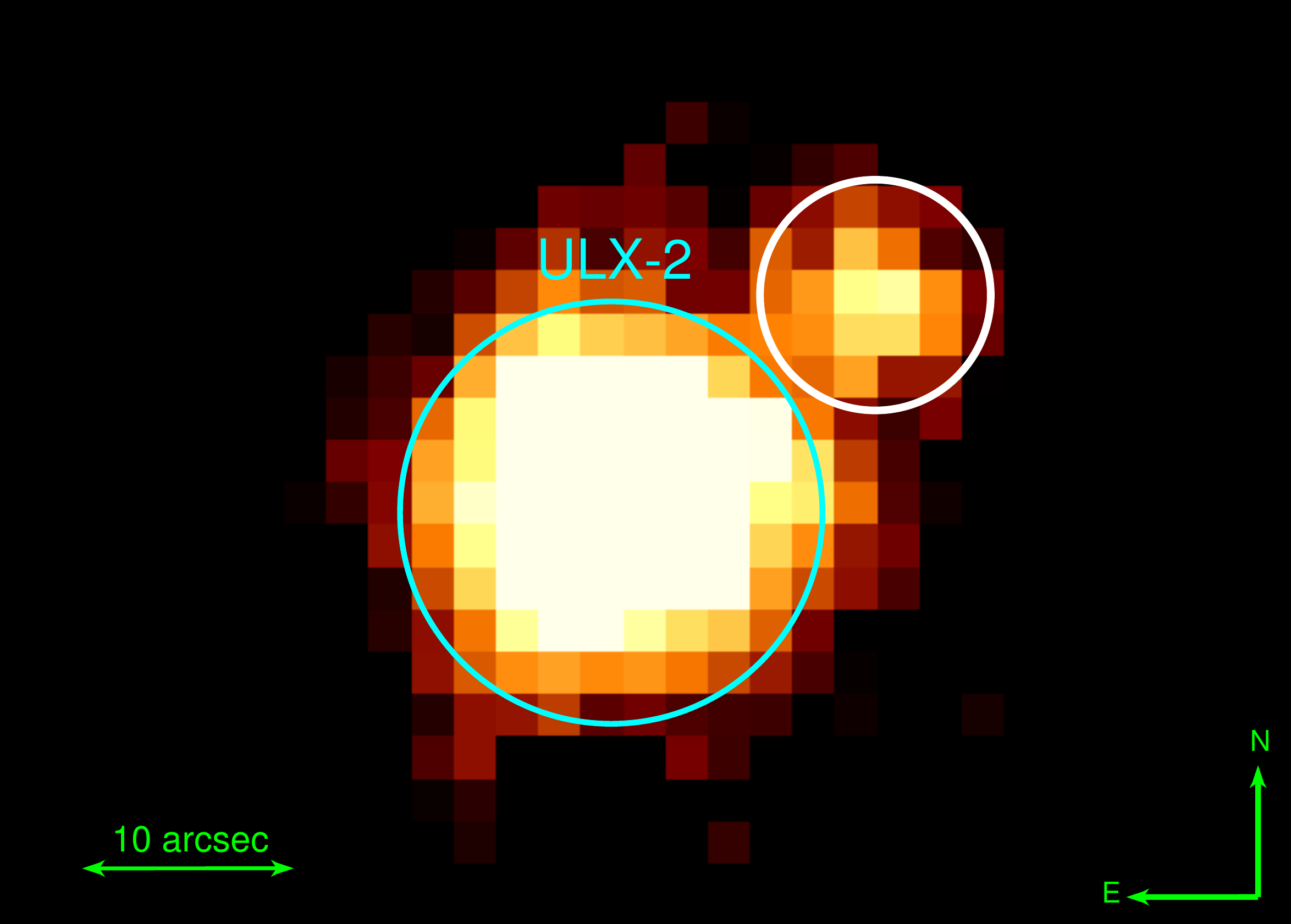}
\caption{2018 EPIC image of the field of ULX-2 (cyan circle). A weaker and soft source (white circle) lies at $\sim15''$ to ULX-2.}  
   \label{zoom_ds9}
   \end{center}
\end{figure}

The deep 2018 \xmm\ observation allowed us to find that ULX-2 lies close to another source at $\sim15''$ (see Figure~\ref{zoom_ds9}). This implies that the contamination from the latter can only be partially removed.
Through a Maximum Likelihood analysis based on the EPIC-pn/MOS point spread functions \citep[see the approach in][]{rigoselli18}, we extracted the spectrum of the second source and we estimated that it is very soft (most of the photons are below 2 keV). It can be modelled ($\chi^2_{\nu}<1$) by a single absorbed power-law with $\Gamma=3.3_{-1.2}^{+2.4}$ and nH $=(1.9_{-1.9}^{+0.4})\times10^{21}$ cm$^{-2}$. We estimated an absorbed 0.3--10 keV flux of $(8_{-3}^{+5})\times10^{-15}$ erg cm$^{-2}$ s$^{-1}$. Assuming the source is in NGC 7456, this corresponds to a luminosity of $\sim2\times10^{38}$ erg s$^{-1}$, which is well below the ULX luminosity threshold and consistent with the Eddington limit of a NS. 
We included the spectral model of the contaminants in the fit of the ULX-2 spectra.

\subsubsection{Temporal analysis}
In the 2005 observation, ULX-2 was the second most luminous ULX in the galaxy and apparently at a constant flux level during the observation. The 2018 source light-curve showed instead a limited short-term variability, over-imposed to a general decay trend during the observation (Figure~\ref{lc_ulx}-center). We modelled the lightcurve with a constant and it does not fit the data well (Null Hypothesis Probability $=1.1\times10^{-5}$).
No coherent pulsations were detected with a $3\sigma$ upper limit of $17-19\%$ for periods in the range 150ms--500s. This is similar to ULX-1 because they have comparable statistics.

\subsubsection{Spectral analysis}
We fitted the 2018 average EPIC-pn and MOS spectra of ULX-2 with an absorbed {\sc diskbb+bbbodyrad} model, that provided a very good fit ($\chi^2=332.8$ for 311 dof; see Table~\ref{spec_par}). The emitting radii for the two components were 400--1400 km (assuming an inclination angle between 5$^\circ$--85$^\circ$ and no color-correction factor) and $102_{-40}^{+45}$ km. 
We estimated an unabsorbed 0.3--10 keV flux of $(2.8\pm0.1)\times10^{-13}$ erg cm$^{-2}$ s$^{-1}$, corresponding to a luminosity of $(8.4\pm0.2)\times10^{39}$ erg s$^{-1}$. Hence this was the most luminous ULX in the galaxy during the observation.

We also tested a thick disc scenario, fitting the spectra with a single {\sc diskpbb} model and finding that it fits the data very well ($\chi^2/\text{dof}=324.41/312$). The model shows some evidence for a thick disc ($p\sim0.5$) with inner temperature of $\sim3$ keV (see Table~\ref{spec_par}). We also added a {\sc diskbb} to the {\sc diskpbb} model, to describe the low energy part of the spectrum. The improvement given by its inclusion was however not statistically significant ($\Delta\chi^2\sim3$, for 2 additional dof).

In 2005, the signal-to-noise ratio was poor ($\sim$940 net counts) and the data were fully affected by high background. Fits of a single {\sc diskbb} or power-law can provide good results (see Table~\ref{spec_par}). 
We tried also to fit the spectrum with the 2018 best-fit model, leaving only the {\sc diskbb} and {\sc bbody} normalizations free to vary. This model was, as expected, acceptable with best-fit normalizations consistent with the previous values. From them, we estimated an unabsorbed 0.3--10 keV flux of $(3.0\pm1.0)\times10^{-13}$ erg cm$^{-2}$ s$^{-1}$ and a luminosity of $(9\pm3)\times10^{39}$ erg s$^{-1}$.

\subsection{ULX-3 and ULX-4}

ULX-3 and ULX-4 were detected only in the 2005 observation with a luminosity of $\sim9.2\times10^{38}$ erg s$^{-1}$ and $\sim1.2\times10^{39}$ erg s$^{-1}$, respectively \citep{walton11}. In the 2018 observation, they were below the detection threshold and we estimated a $3\sigma$ upper limit on their 0.3--10 keV flux of $3.7\times10^{-15}$ erg cm$^{-2}$ s$^{-1}$, corresponding to a luminosity of $\sim1.3\times10^{38}$ erg s$^{-1}$, i.e. well below the ULX threshold, and implying a factor $\geq$10 variability on long timescales.

\subsection{ULX-5}

During the 2005 observation, ULX-5 was undetectable and we estimated a $3\sigma$ upper limit on the 0.3--10 keV flux of $3.1\times10^{-14}$ erg cm$^{-2}$ s$^{-1}$, corresponding to a luminosity of $\sim1.1\times10^{39}$ erg s$^{-1}$. The 2018 source light-curve is shown in Figure~\ref{lc_ulx}-right and no significant intra-observation variability was observed. Its spectrum was hard and could be modelled with a single absorbed power-law, with photon index of 1.9 (Figure~\ref{spec_ulx}-right and Table~\ref{spec_par}). The measured column density is higher than the Galactic one.
The 0.3--10 keV luminosity was $(1.3\pm0.1)\times10^{39}$ erg s$^{-1}$. 

The hard spectrum and lack of intra-observation variability make the association with a foreground flaring star unlikely \citep[e.g.][]{pye15}. Furthermore, in ESO/EFOSC2 observations in the B-band (5$\times$ 180 s) and in the R-band (5 $\times$ 300 s) we did not find any optical counterpart down to B$\sim$22 mag and R$\sim$23 mag. However, WISE detected within 2'' a source with W1 magnitude of $17.4 \pm 0.2$ mag. 

A Galactic X-ray source, assuming a reasonable distance of about 5\,kpc, would have had a 0.3--10 keV luminosity of $\sim2\times10^{32}$ erg s$^{-1}$. Should it be a magnetar, such a luminosity would be expected only in late stages of an outburst \citep[see e.g][]{rea11,cotizelati18}. However, the sky position of the source, with a Galactic latitude $l\sim-34$\textdegree, is not likely, given the spatial distribution of magnetars in our Galaxy \citep{olausen14}. A long-lasting Galactic accreting binary system outburst is unlikely as well, because we would expect to observe an optical counterpart (not detected). On the other hand, a low mass X-ray binary in quiescence cannot be excluded \citep[e.g.][]{plotkin13}.

Because of its position with respect to the galaxy and its hard spectrum, we cannot rule-out that it is a background active galactic nucleus (AGN). In fact, the upper limit to the optical-to-X-ray flux ratio of $\leq0.8$ is consistent with those typical for AGNs and Blazars \citep[see e.g.][]{maccacaro88}. Furthermore, from the Log\,$N$--Log\,$S$ of extragalactic sources \citep{moretti03}, we could estimate that $\sim100$ objects are expected to be found in a square degree at the observed flux of ULX-5. Hence, for a galaxy dimension of $5.01'\times1.78'$, we estimated $\sim0.2$ background AGN in the field.
Although we propose this source as a likely ULX candidate located in the outskirts of NGC 7456, deeper multi-wavelength observations are needed to test its ULX nature and rule-out a background AGN. 

\section{Discussion}

In this work, we have carried out the first, detailed investigation of the properties of several ULXs hosted in the galaxy NGC 7456, thanks to a long-exposure \xmm\ observation taken in 2018 { (the only X-ray observation apart for a much shorter \xmm\ observation taken in 2005)}. Two sources, ULX-1 and ULX-2, are very bright with luminosities close to $10^{40}$ erg s$^{-1}$ and were active during the two \xmm\ observations. On the other hand, two ULXs in the galaxy (ULX-3 and ULX-4) may be considered as candidate transient objects because they were not detected anymore in 2018 (although further monitoring is necessary to confirm their nature). Moreover, a possible new ULX candidate, with luminosity slightly higher than $10^{39}$ erg s$^{-1}$, was discovered in the 2018 data. 

{ Unfortunately, no {\it HST} observations of the galaxy at the position of the ULXs have been taken, therefore the optical data are of quite poor resolution and little can be inferred about the ULX counterparts. We could associate ULX-1, ULX-2 and ULX-4 to the spiral arms of NGC 7456, while ULX-3 is close to the galactic center. We searched multi-wavelength archives and we found only {\it WISE} and {\it Galex} observations: their resolution is however not adequate to distinguish the emission from the single ULX counterparts from the faint diffuse, unresolved emission from the galaxy. For these reasons, no evidence of bubbles (or nebulae) can be inferred. Deeper observations are necessary to investigate better the nature of the ULX counterparts.}

\subsection{ULX-1}

ULX-1 is a bright source (already reported in \citealt{walton11}) located in a spiral arm, at $\sim2.3$' from the center of NGC 7456. The archival optical/infrared images are not deep enough to allow us to find evidence of a clear counterpart. We only estimated a 5$\sigma$ upper limit on the magnitudes $M_J\geq20.2$ and $M_{k_s}\geq18.1$ mag (from the Vista Hemisphere Survey; \citealt{mcmahon13}). Our analysis showed that ULX-1 is characterized by pronounced short-term variability on time scales down to 500s.

The 0.3--10 keV source luminosity varied across the 2005 and 2018 observations in the range $(6-10)\times10^{39}$ erg s$^{-1}$. The high quality 2018 ULX-1 spectrum showed that the source was quite soft (the ratio of the 1--10 keV and 0.3--1 keV fluxes was $\sim30\%$) and we found that its spectral shape can be well modelled with the combination of two thermal components. Because of its spectral and temporal properties, we can claim that the source was not in the typical {\it hard} or {\it soft} states of the Galactic accreting BHs in outburst \citep[e.g.][]{mcclintock06,belloni11}, hence we can consider less likely the scenario of sub-Eddington accretion onto a massive BH of 40--100 M$_{\odot}$. 

If the source is accreting at super-Eddington rates, the compact object might be surrounded by an accretion disc, or an extended outflow, with an inner radius of 2000--7000 km. Outflows are expected in the case of super-Eddington accretion \citep{poutanen07}. These may be identified with the discovery of blueshifted absorption and emission lines (e.g. \citealt{pinto16,kosec18}) or X-ray bubble nebulae around ULXs \citep{belfiore19}. In our analysis, we report on a possible absorption feature at $\sim0.7$ keV which is often observed in ULX spectra \citep[e.g.][]{middleton15b}. The feature may be then associated to a blending of ionized Oxygen features (O VII-O VIII). 

However, the line might also be interpreted as a cyclotron line in an accreting NS scenario, as for the source M51 ULX-8 that has been proposed as a candidate NS because of the first detection of a cyclotron absorption feature \citep{brightman18,middleton19}.
The feature in ULX-1 might correspond to a NS magnetic field of $B_{12} = (1+z) E_{c}/11.6\text{ keV}\sim8\times10^{11}$ G and $B_{15} = (1+z) E_c/6.3\text{ keV} \sim 1.5\times 10^{14}$ G for an electron or proton feature, respectively, where $z$ is the gravitational redshift assumed to be 0.3 for a NS, $E_c$ is the line energy (0.67 keV), and $B_{12}$ and $B_{15}$ are the magnetic fields in units of $10^{12}$ G and $10^{15}$ G, respectively. 

According to \citet{sutton13}, ULX-1 can be  classified as a {\it soft-ultraluminous} source. It has been shown that the sources of this class are generally those with the highest short-term flux variability. This is confirmed by the 2018 data of ULX-1, in which there was a strong flux variability described by an overall flux increment imposed on short-term variability on timescales from hundreds to thousands of seconds ($F_{var}\sim50\%$). This is amongst the highest variability ever observed in a ULX \citep[e.g.][]{sutton13}. The source lightcurve is characterised by flux variations of more than an order of magnitude in $\sim80$ ks with indications of possible flares or dips.

The flux evolution is also accompanied by spectral changes, although the interpretation is not straightforward. 
We found that the variability is mainly driven by the hard component as already reported for the variability of soft-ultraluminous ULXs  \citep[e.g.][]{middleton15a}.
The time evolution of the hardness ratios suggests that the spectral softening detected during low flux periods might be due to a reduction of the hard component emission. Such a result rules out a dipping activity caused by photoelectric absorption, as in such a case a hardening rather than a softening of the X-ray spectrum would be expected.
However, the ULX-1 properties are similar to those observed in NGC 6946 ULX-3 \citep{earnshaw19b}, NGC 253 ULX-1 \citep{barnard10}, NGC 5907 ULX-1 and NGC 55 ULX-1 (although this is markedly fainter - $\sim10^{39}$ erg s$^{-1}$ - and shows well defined dips on time-scales of hundreds of seconds; e.g. \citealt{stobbart04}), which are all soft-ultraluminous sources. In particular, NGC 5907 ULX-1 and NGC 55 ULX-1 are characterized by very soft spectra, where blueshifted ($\sim0.2$c) absorbtion features are observed and associated to powerful outflows that obscure the inner and hotter regions of the accreting system \citep[e.g.][]{pinto17}.

It has been proposed that the short-term variability observed in the soft ULX can be ascribed to the turbulence of the outflows that intersect, from time to time, the line of sight (LoS). Hence this leads to the conclusion that the sources are seen from a high inclination angle. We may therefore propose that also ULX-1 is inclined so much that our LoS is close to the border of the outflow photosphere, which can randomly obscure/unveil the inner, hotter regions close to the compact object as in NGC 55 ULX-1. Alternatively, the variability may be ascribed to an unstable mass transfer rate. 
Furthermore, ULX-1 shows a variability which may resemble some states of the Galactic BH binary GRS 1915+105, such as the $\Theta$ or $\rho$ classes \citep[e.g.][]{belloni00}, which however happen on shorter time scales (the data quality is of course much higher than that of ULX-1). GRS 1915+105 is a well known swinging Eddington/super-Eddington accreting source \citep[e.g.][]{vilhu99} and its complex variability was also ascribed to a variable wind \citep[e.g.][]{neilsen12}. 

Furthermore, the high short-term variability of ULX-1 is similar to that observed in M51 ULX-7 (\citealt{liu02,earnshaw16a,rodriguez19}), which is a PULX and has variability of a 30-40$\%$ on timescales of a few ks as well. This might be an additional hint that ULX-1 hosts a NS in which pulsations are undetected because of the high inclination viewing angle. On the other hand, it is important to remark that M51 ULX-7 is spectrally hard, hence different from ULX-1.
However, should the ULX-1 compact object be a NS with a spin period of $\sim$1s (as seen in most PULXs), the observed source flux variability could not be ascribed to propeller effects as it would produce a bimodal luminosity variation of more than two orders of magnitude: $\Delta L\sim170P^{2/3}M^{1/3}_{1.4 M_{\odot}}R^{-1}_{10^6 cm} \sim 170$ \citep[e.g.][]{corbet96,campana01,mushtukov15,tsygankov16,campana18}. On the other hand, a period smaller than 1s can produce the observed luminosity jump.

Finally, \citet{earnshaw19} showed that among a population of $\sim300$ ULXs, only 8 sources were found to be highly variable (RMS$>$30$\%$) on timescales of hundreds to a few thousand ks. Using the {\it XMM-DR4} catalogue, we found that if we limit our investigation to the ULXs with observations longer than 30 ks, we have 5--6 variable objects for a total of 182 selected sources. Furthermore, if we consider only the sources with count-rates higher than 0.1 cts s$^{-1}$, the number of variable ULXs is 4 on a total of 23 selected sources.
This implies that casual observing might lead to some bias in the detection of periods of high amplitude variability.

\subsection{ULX-2}

The source was detected in both \xmm\ observations at a luminosity of $8-9\times10^{39}$ erg s$^{-1}$, with a low short-term flux variability (apart for a slight decay in the 2018 data). 

Our spectral analysis showed that a phenomenological model given by two thermal components well describes its spectrum. 
Unfortunately, the quality of the 2005 data did not allow us to constrain any spectral variability.
From the higher quality 2018 data, we estimated temperatures of 0.6 keV and 1.3 keV associated to emitting radii of 400--1400 km and $\sim100$ km, for the {\sc diskbb} and {\sc bbodyrad} model, respectively. However, we cannot exclude a super-Eddington scenario in which accretion is driven by an advection dominated disc with a temperature of $\sim3$ keV, as observed in other ULXs \citep[e.g. IC 342 X-1, NGC 5643 ULX-1;][]{gladstone09,pintore16}.

Because of its hard spectral shape, the source may be classified as a {\it hard-ultraluminous} or a {\it broadened disc} ULX \citep{sutton13}. In a super-Eddington scenario, the source is likely observed from a low-inclination angle, where the line of sight enters into the funnel of the outflows \citep[e.g.][]{middleton15a}. The non detection of pulsations does not allow us to establish if the compact object is a NS or a BH. Simultaneous broadband observations with \xmm\ and {\it NuSTAR} will permit to better constrain the source nature, for example testing the existence of a third, high-energy spectral component, usually observed in the PULX spectra and likely arising from column accretion \citep[e.g.][]{walton18} on top of the NSs.

\subsection{Variable ULXs}
In \citet{walton11}, ULX-3 and ULX-4 were only marginally classified as ULXs since their luminosity was $\sim10^{39}$ erg s$^{-1}$.
On the other hand, the two sources are not detected in our new observation down to a limit of $<2\times10^{38}$ erg s$^{-1}$, making them candidate transient ULXs (tULXs). 
By now, only a dozen ULXs are known to be transient; however casual observational scheduling might have some bias in the observed variability pattern and the detection of periods of high variability in ULXs. Furthermore, most of the tULXs have been discovered only by chance \citep[e.g.][]{soria12,esposito13,middleton13,carpano18, pintore18a, vanhaften19, earnshaw18}. Therefore, the actual fraction of tULXs and their duty cycle are poorly constrained. Regular high quality X-ray monitoring of a large sample of galaxies hosting ULXs (hence seemingly having a suitable environment for the production of ULXs) are needed.

We stress that all confirmed PULXs belong to the transient group, and the switch-off may be caused by propeller effects \citep[e.g.][]{israel16a,tsygankov16} which will cause a drop in luminosity of $\Delta L\sim 170P^{2/3}M^{1/3}_{1.4 M_{\odot}}R^{-1}_{10^6 \rm cm}$. Assuming a spin period $\leq$1s, the $3\sigma$ flux upper limits are compatible with the entrance in propeller for both ULX-3 and ULX-4.
Therefore, it is possible that, even though the pulsations are not detected (perhaps due to limited statistics), ULX-3 and ULX-4 could potentially host NSs. We cannot rule-out an alternative explanation for the high level of variability amongst observation being due to high amplitude super-orbital modulations. Furthermore, a possibility is that these ULXs host BHs and they were observed during a particularly bright outburst \citep[e.g.][]{esposito13, middleton13, earnshaw18}

Finally, similar considerations also apply to the new source (ULX-5), should it be confirmed as a genuine ULX in the NGC 7456 galaxy.

\section{Conclusions}

In this work we have presented a full study of the ULX population in the galaxy NGC 7456. Two of these are bright sources with luminosities of $5\times10^{39}-10^{40}$ erg s$^{-1}$ (which might be persistent sources) and spectrally described by a two-component thermal model as usually seen in ULX spectra. However, at least one of the two (ULX-1) is highly variable (varying by an order of magnitude) on timescales of hundreds of seconds to ks, presenting one of the largest flux variations ever observed in a ULX ($\sim50\%$ fractional variability). Such a variability is mainly driven by the high-energy part of the emission. The nature of its compact object is not yet clear. We propose that the source is seen with an inclination angle such that our line of sight occasionally straddles the optically thick turbulence of an outflow, which covers from time to time the inner regions where the high-energy emission is produced. Alternatively, the variability may be related to changes in the accretion rate. ULX-1 increases the sample of highly variable sources and we cannot exclude that the compact object is a NS.

ULX-2 is instead spectrally rather hard and can also be modelled by a single thick disc, suggesting a super-Eddington accreting scenario.

Two other ULXs (ULX-3 and ULX-4) in the galaxy are variable, and seem to reach and marginally overcome the ULX threshold. These can be considered as transient ULX candidates. Finally, we found a possible new bright source (ULX-5) in the galaxy, the nature of which is not yet constrained. Further observations will be necessary to determine the possible transient nature of these sources.

\section{acknowledgement}

The scientific results reported in this article are based on 
observations obtained with \xmm, an ESA science mission with 
instruments and contributions directly funded by ESA Member States and NASA.
LS acknowledges support from the research grant “iPeska” (P.I. Andrea Possenti) funded under the INAF national call Prin-SKA/CTA approved with the Presidential Decree 70/2016.
GR acknowledges the support of high performance computing resources awarded by CINECA (MARCONI and GALILEO), under the ISCRA initiative and the INAF-CIENCA MoU; and also the computing centres of INAF -- Osservatorio Astronomico di Trieste and Osservatorio Astrofisico di Catania, under the coordination of the CHIPP project, for the availability of computing resources and support. CP is supported by ESA Research Fellowships. DJW and MM acknowledges support from STFC in the form of an Ernest Rutherford Fellowship. HPE acknowledges support under NASA contract NNG08FD60C. FB is funded by the European Union’s Horizon 2020 research and innovation programme under the Marie Sk\l odowska Curie grant agreement no. 664931. This research was supported by high
performance computing  resources at New York University Abu Dhabi. HPE acknowledges support under NASA contract NNG08FD60C. TPR acknowledges support from STFC as part of the consolidated grant ST/K000861/1. A.P. acknowledges financial support from the Italian Space Agency and National Institute for Astrophysics, ASI/INAF, under agreements ASI-INAF I/037/12/0 and ASI-INAF n.2017-14-H.0.

We acknowledge funding in the framework of the project ULTraS ASI--INAF contract N.\,2017-14-H.0, and project ASI--INAF contract I/037/12/0.

\addcontentsline{toc}{section}{Bibliography}
\bibliographystyle{mn2e}
\bibliography{biblio}

\label{lastpage}
\end{document}